%% file: ms.tex
\documentclass[10pt,twocolumn,letterpaper]{article}

\pdfoutput=1

\usepackage{arxiv}
\usepackage{booktabs} 
\usepackage[ruled,linesnumbered]{algorithm2e} 
\usepackage{graphicx}
\usepackage{amsmath}
\usepackage{bbm}
\usepackage{diagbox}
\usepackage{wrapfig}
\usepackage{sidecap}

\arxivfinalcopy 
\def\arxivPaperID{0000} 
\ifarxivfinal\fi

\begin{document}

\title{Enabling Simulation of High-Dimensional Micro-Macro Biophysical Models through Hybrid CPU and Multi-GPU Parallelism}

\author{Steven Cook\\
University of California, Riverside\\
{\tt\small scook005@cs.ucr.edu}
\and
Tamar Shinar\\
University of California, Riverside\\
{\tt\small shinar@cs.ucr.edu}
}

\maketitle

\begin{abstract}
Micro-macro models provide a powerful tool to study the relationship between microscale mechanisms and emergent macroscopic behavior. However, the detailed microscopic modeling may require tracking and evolving a high-dimensional configuration space at high computational cost. In this work, we present a parallel algorithm for simulation a high-dimensional micro-macro model of a gliding motility assay.  We utilize a holistic approach aligning the data residency and simulation scales with the hybrid CPU and multi-GPU hardware.  With a combination of algorithmic modifications, GPU optimizations, and scaling to multiple GPUs, we achieve speedup factors of up to 27 over our previous hybrid CPU-GPU implementation and up to 540 over our single-threaded implementation.  This approach enables micro-macro simulations of higher complexity and resolution than would otherwise be feasible.
\end{abstract}

\input{ms_body}

\bibliographystyle{ieee}
\bibliography{BibMT} 

\end{document}

%% file: ms_body.tex
\input{commands}
\section{Introduction}

Active gels exhibit macroscopic flow structures driven by the detailed microscopic interactions of constituent elements. Pronuclear centering and migration and cytoplasmic streaming are two such examples, both being critical cellular processes driven by filament-motor mixtures. Reduced-component studies have found these systems to be highly sensitive to the microscopic interactions between motors and filaments; for instance, the detachment time of a motor protein at a filament end affects whether filaments form networks of asters or vortices \cite{N2002,NS2001,NF2007}. Additionally, the tens-of-nanometers sized motor proteins bind, walk along, and detach from micrometer-length filaments on a faster timescale than the filament network evolution. Simulating even a millimeter-sized system with such disparate length and time scales and sensitivity to detailed interactions thus poses a challenging computational problem. Tracking interacting Lagrangian particles can become infeasible with large quantities of microstructural elements.

A promising approach lies in micro-macro methods, which couple a kinetic theory model of the microstructure (here, the configuration of the motors and filaments in the active gel) to the macroscale continuum mechanical representation of a viscoelastic fluid \cite{keunings2004micro}. Kinetic theory models have been applied in the study of biological active matter \cite{HS2011}, self-propelled particles \cite{ss2008b}, and networks of neurons \cite{cai2004effective}. They enable detailed microscale modeling that would otherwise be lost via closure approximations in macroscopic modeling approaches, and are particularly useful at scales where tracking individual particles and their interactions would be prohibitive. Compared to purely macroscopic methods, micro-macro methods are more computationally demanding, as they require evolving the microstructure density in a potentially high-dimensional configuration space. 

\cite{HCS2014} and \cite{CHS2017} developed a micro-macro model for a gliding motility assay, consisting of immersed rigid filaments that glide along motor proteins anchored to the substrate of a chamber immersed in viscous fluid. This model includes hydrodynamic and steric interactions between the filaments. A high-dimensional kinetic theory describes the evolution of the filaments and motors. To make this model computationally feasible, parts of the microscale computation were ported to the GPU using Nvidia's CUDA C language \cite{nvidia2011nvidia}. In this work, we enable faster and significantly more detailed simulations through holistic restructuring of this algorithm, aligning the computation and data flow with the underlying heterogenous computational resources. Moreover, these changes facilitate scaling to multiple GPUs across separate machines with MPI. We further utilize a variety of CPU and GPU optimizations.
Our work expands the range of micro-macro models which can be simulated by direct solution of the kinetic theory and coupling equations to models with higher dimensional configuration spaces, at higher resolutions \cite{keunings2004micro}. To our knowledge, \cite{HCS2014} is the first GPU-accelerated micro-macro kinetic theory-based simulation. \cite{HCS2014} achieved up to ~20x speedups over a single-threaded CPU implementation, while the algorithm presented here achieves a further 27x speedup over \cite{HCS2014} and \cite{CHS2017}. Key to our approach is moving the microscopic scale and related tasks, which are smaller scale in both space and time, to the GPU and limiting CPU-GPU communications to the longer timescale of the filaments and fluid. Such holistic approaches are recommended to achieve scalability in heterogeneous environments \cite{niemeyer2014recent}, \cite{ge2013multi}. We note that our method does not suffer from common GPU simulation challenges encountered in various other approaches such as building adjacency lists \cite{jiang2016computational}, reordering storage based on cell location \cite{westphal2014multiparticle}, dynamic, irregular data accesses \cite{zabelok2015adaptive}, thread divergence \cite{frezzotti2011solving}, or neighbor exchanges of halo regions. 

The paper is organized as follows. A description of the model and implementation is presented in Section~\ref{sec:model}, algorithmic, data flow, GPU, and MPI modifications are discussed in Section~\ref{sec:optimizations}, results are discussed in Section~\ref{sec:results}, and we conclude in Section~\ref{sec:conclusions}.

\section{Motility Assay model}
\label{sec:model}

\begin{figure}
\centering
\includegraphics[scale=.4]{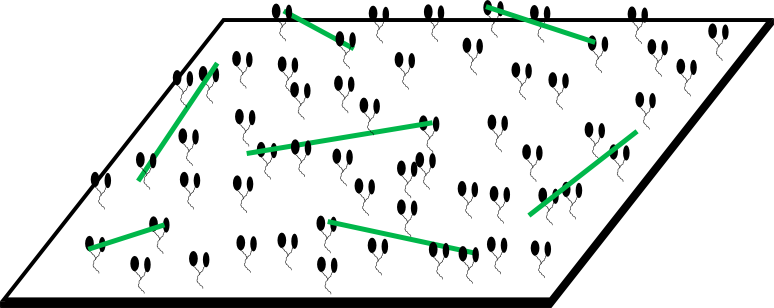}
\caption{Gliding motility assay. Motor proteins (black) anchored to the substrate bind to filaments (green), walk along them and exert forces, then detach.}
\label{fig:assay}
\end{figure}

Figure~\ref{fig:assay} illustrates a gliding motility assay. The geometry consists of top and bottom plates separated by a narrow vertical gap, which is filled with a viscous fluid containing ATP fuel. At the bottom plate, motor protein tails are anchored to a substrate. The motor protein heads diffuse in solution, tethered to their tail by a flexible stalk. When a filament enters the capture radius of a motor protein head, the head may bind to the filament. As the bound motor head walks toward the filament plus end, it exerts force, causing the filament to glide in the opposite direction, until the motor head detaches.  When many filaments are present, the underlying microscopic mechanism coupled with hydrodynamic and steric interactions give rise to a variety of emergent macroscopic behaviors such as a lattice of vortices \cite{SNSTYCO2012}.  Through our modeling and simulation, we aim to better understand the relationship between the microscale interactions and the macroscopic phenomena.  Related problems of emergent self-organization from simple interactions include flocking and swarming of birds, fish, and bacteria.

The model equations are presented in simplified, nondimensionalized form in Table~\ref{table:equation_summary}. The filament density is parameterized by center-of-mass location $\x$, orientation $\p$, and time $t$ as $\Psi(\x,\p,t)$. 

The configuration space of bound motors is higher dimensional, as we need to track the center-of-mass position $\x$ and orientation $\p$ of the filament a bound motor with tail anchored at $\r_0$ is bound to, along with its arclength parameter along that filament $s$.
\begin{wrapfigure}{r}{.15\textwidth}
\begin{center}
\hspace{-.25in}
\includegraphics[width=.15\textwidth]{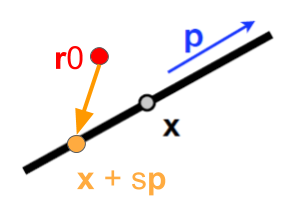}
\vspace{-.125in}
\end{center}
\end{wrapfigure}
This yields the high-dimensional density $\M_b(\r_0,\x,\p,s,t)$ of bound motors per filament, illustrated in the figure to the right. 
A key observation is that motors with tail anchored at position $\r_0$ on the assay substrate can only bind to filament sections that are within the capture radius of the motor stalk $r_c$. This greatly reduces the feasible configurations a motor protein head may be bound in, and obviates the need to track configurations $|\x+s\p- \r_0| > r_c$. We denote by $B_{rc}(\r_0)$ all feasible $\x,\p,s$ configurations such that $|\x+s\p- \r_0| \le r_c$. We do not track unbound (free) motor heads, only their tail position $\r_0$, so the density of free motors $\M_f(\r_0,t)$ is two-dimensional. We model the filament and bound motor protein densities as distributed by a smooth Dirac delta function in $z$ about a plane a small distance $z_0$ above the bottom plate, i.e., $\Psi(\x,\ldots) = \Psi_{z_0}(\x_2,\ldots) \delta(z)$ and 
$\M_{b}(\r_0, \x,\cdots) = \M_{b,z_0}(\r_0, \x_2,\cdots) \delta(z)$
. We thus evolve the lower-dimensional $\M_{b,z_0}$ and $\Psi_{z_0}$ in our simulation. We drop the $z_0$ from $\Psi$ and $\M_b$ in the remainder of the paper for brevity.

\begin{table}[h]
\begin{tabular}{p{\linewidth}}
\hline
Filament equations
{\begin{align}
&\quad\partial_t\Psi+\nabla_2\cdot(\dot{\x}_2\Psi)+\partial_\theta(\dot\theta\Psi)=0 \label{eq:filament_1}\\ 
\quad\dot{\x}_2&=-V_\text{sp}\p_2+\vec{u}_2+U_{t,\parallel}^0\p_2\p_2:\nabla_2\vec{D}_{2,z_0}\label{eq:filament_2}\\
& \quad\quad\quad\quad -D_{t,\parallel}\nabla_2\ln\Psi \nonumber\\
\quad\dot{\theta}&=\nabla_2\vec{u}_2+U_r^0\vec{D}_{2,z_0}:\p_2^\perp\p_2-D_r\partial_\theta\ln\Psi \label{eq:filament_3}
\end{align} }
Motor equations
{
\begin{align}
 \partial_{t} \M_{b} +\partial_{s} \M_{b} &=  -  k_{\text{off}} \M_{b} +k_{\text{on}}\M_f \mathbbm{1}_{D_{r_c}} \label{eq:motor_1}\\
 \M_{b,coarse} &=\iiint\M_{b}\Psi\, ds\,d\x_2 \,d\theta \label{eq:motor_2}\\
 \M_f &=\M-\M_{b,coarse} \label{eq:motor_3}
\end{align}
}
Fluid equations 
{
\begin{align}
-\nabla_{2}^2 \vec{u}_2-\frac{1}{\varepsilon^2}\partial_{zz}\vec{u}_2+P_0 \nabla_{2}q &= \sigma_f \nabla_{2} \cdot \Sf \label{eq:fluid_1}\\
  & \quad\quad -\sigma_t \nabla_{2} \cdot \St  + \vec{F}_2 \nonumber \\
- \nabla_{2}^2 w -\frac{1}{\varepsilon^2} \partial_{zz}w+P_0\partial_{z}q &= 0 \label{eq:fluid_2}\\
\nabla_2\vec{u}_2+\partial_{z}w &= 0 \label{eq:fluid_3}
\end{align}
}
Motor force 
{\begin{align}
\vec{F}_2(\x_2) &= F  \iiiint \vec{f}(\y_2,\p_2,s) \,ds\,d\vec{r}_0\,d\y_2\,d\theta \label{eq:force_1}\\
\vec{f}(\y_2,\p_2,s) &= \p_2  \delta(\y_2+\frac{l}{L}s\p_2-\x_2) \Psi\M_{b} \nonumber
\end{align}
}\\
\hline
\end{tabular}
\caption{Summary of model equations for the filament and motor protein densities, the macroscopic fluid equations, and the motor force that couples them.}
\label{table:equation_summary}
\end{table}

 We represent the fluid velocity in three dimensions, with periodic boundary conditions in the $x$ and $y$ dimensions and no-slip conditions in the $z$ dimension at the top and bottom plates. The system evolves on two timescales; the motors bind to, walk along, and unbind from the filaments on a faster timescale than the filaments and fluid evolve. Bound motor heads generate forces that are spread onto the fluid in an immersed boundary method fashion \cite{P2003}. Together with stress terms arising from filament inextensibility and steric interactions \cite{ESS2013}, the motor forces (Eq.~\eqref{eq:force_1}) couple the densities $\Psi$ and $\M_b$ to the fluid velocity (Eq.~\eqref{eq:fluid_1}). The $x$ and $y$ dimensions are discretized over a regular square grid, and the $z$ dimension is discretized over an adaptive grid that is finely resolved around $z=z_0$ near the bottom plate and becomes coarser moving toward the upper plate. Allowable filament orientations are constrained to the $(x,y)$ plane, so we can represent $\p=(\cos \theta,\sin \theta, 0 )^T$. Orientation $\theta$ and arclength parameter $s$ are discretized uniformly with the same resolution.

Algorithm~\ref{alg:full_simulation} summarizes the process for evolving the filaments, motor proteins, and fluid velocity as in \cite{HCS2014}, \cite{CHS2017}. First, we compute the adaptive time steps based on their stability conditions, with outer time step $dt$ restricted by the advective fluxes in Eq.~\eqref{eq:filament_1}, and the inner time step $dt^*$ restricted by the motor speed and binding/unbinding rates in Eq.~\eqref{eq:motor_1}.
Next, $\Psi(t+dt)$ is solved using Crank-Nicolson for the diffusive terms and Adams-Bashforth 2 with upwinding for the advective terms (line 4). The bound motor density evolution routine in lines 7-9 performs the motor protein advection (bound motor heads walking along the filaments toward their plus-ends) and applies a Superbee flux limiter, as well as simulates the binding and unbinding of free and bound motor proteins respectively. After every configuration of $\M_b(\r_0\x,\theta,s,t)$ for a particular $\r_0$ has been updated, the integral $\M_{b,coarse}(\r_0,t)=\iiint \M_b(\r0,\x,\theta,s,t) \Psi(\x,\theta,t) d\x d\theta ds$ is calculated at the same $\r_0$ to ensure that the number of bound motors does not exceed the total number of motors at $\r_0$. If so, all $\M_b$ configurations with that $\r_0$ are scaled down to conserve the total number of motors before $\M_f$ is calculated. We next update the free motor density in line 9.  The extra stress terms $\Sf, \St$ arising from filament inextensibility and steric interactions \cite{CHS2017} are computed as moments of $\Psi$ in line 13. We perform a two-dimensional FFT and solve the transformed system of fluid equations for the three velocity components and pressure $\hat{u},\hat{v},\hat{w},\hat{q}$ at every position on the $z$ grid, then perform an inverse FFT to obtain the three-dimensional fluid velocity $\vec{u}$ in line 13.

In a single-threaded implementation, the high dimensionality of $\M_b$ makes the computations in lines 7-9 and line 11 prohibitively expensive for even moderate grid resolutions and experiment times.  Fortunately, $\M_b$ can be computed explicitly and easily parallelized over $\r_0$.  Thus in \cite{HCS2014,CHS2017}, the $\M_b$ and $\vec{F}$ computations are ported to the GPU.  On the other hand, the $\Psi$ equation is stiff due to the diffusion terms and is computed semi-implicitly on the CPU.  This decomposition of work is similar to several hybrid reactive flow solvers \cite{niemeyer2014recent}.  

To perform the integral in line 11, \cite{HCS2014,CHS2017} use independent GPU threads to compute the integral at each $\r_0$ accumulating the partial results in thread-local storage, limiting the use of atomic operations to the final reduction over nearby $\r_0$ at each $\x$. Extra stress tensor calculation, forward and reverse fast Fourier transforms, and computation of the independent semi-spectral systems are all multithreaded on the CPU. In this work, we expand upon this hybrid computational approach as described below.

\begin{algorithm}[t]
  \caption{Evolution scheme for the coupled microtubule density, motor protein density, and fluid velocity equations.}
  \label{alg:full_simulation}
    {Initialize $\Psi$ and $\M_b$}\\
    \While{$t < t_{\text{end}}$}
    {
      {Compute adaptive time steps $dt,dt^*$}\\
      {Compute filament density $\Psi(t+dt)$ (Eqs.~\eqref{eq:filament_1}-\eqref{eq:filament_3})}\\
      {set $t^*_{\text{end}}=t+dt$}\\
      \While{$t^* < t^*_{\text{end}}$}
      {
        {Compute bound motor density $\M_b(t^*+dt^*)$ (Eq.~\eqref{eq:motor_1})}\\
        {Update coarsened density $M_{b,coarse}$ (Eq.~\eqref{eq:motor_2})}\\
        {Update free motor density $\M_f$ (Eq.~\eqref{eq:motor_3})}\\
      }
      {Calculate motor force $\vec{F}_2$ (Eq.~\eqref{eq:force_1})}\\
      {Calculate extra stresses $\Sf,\St$ (Eq.~\eqref{eq:fluid_1})}\\
      {Calculate fluid velocity $\vec{u}$ (Eqs.~\eqref{eq:fluid_1}-\eqref{eq:fluid_3})}\\
    }
\end{algorithm}

\section{Acceleration Methodology}
\label{sec:optimizations}

Our primary focus in this work is significant performance improvement through targeted algorithmic design enabling a multi-GPU decomposition, as well \todo{as single-GPU optimizations}, described in detail below.
Through these efforts we are able to scale to higher resolutions than previously possible and achieve up to 27x total simulation acceleration in a four GPU configuration. Use of additional GPUs is supported and should provide further speedup with similarly excellent scaling, although we did not test this in this work.

\subsection{Holistic Algorithmic and Data Flow Restructuring}

\begin{figure*}
\centering
\includegraphics[scale=.26]{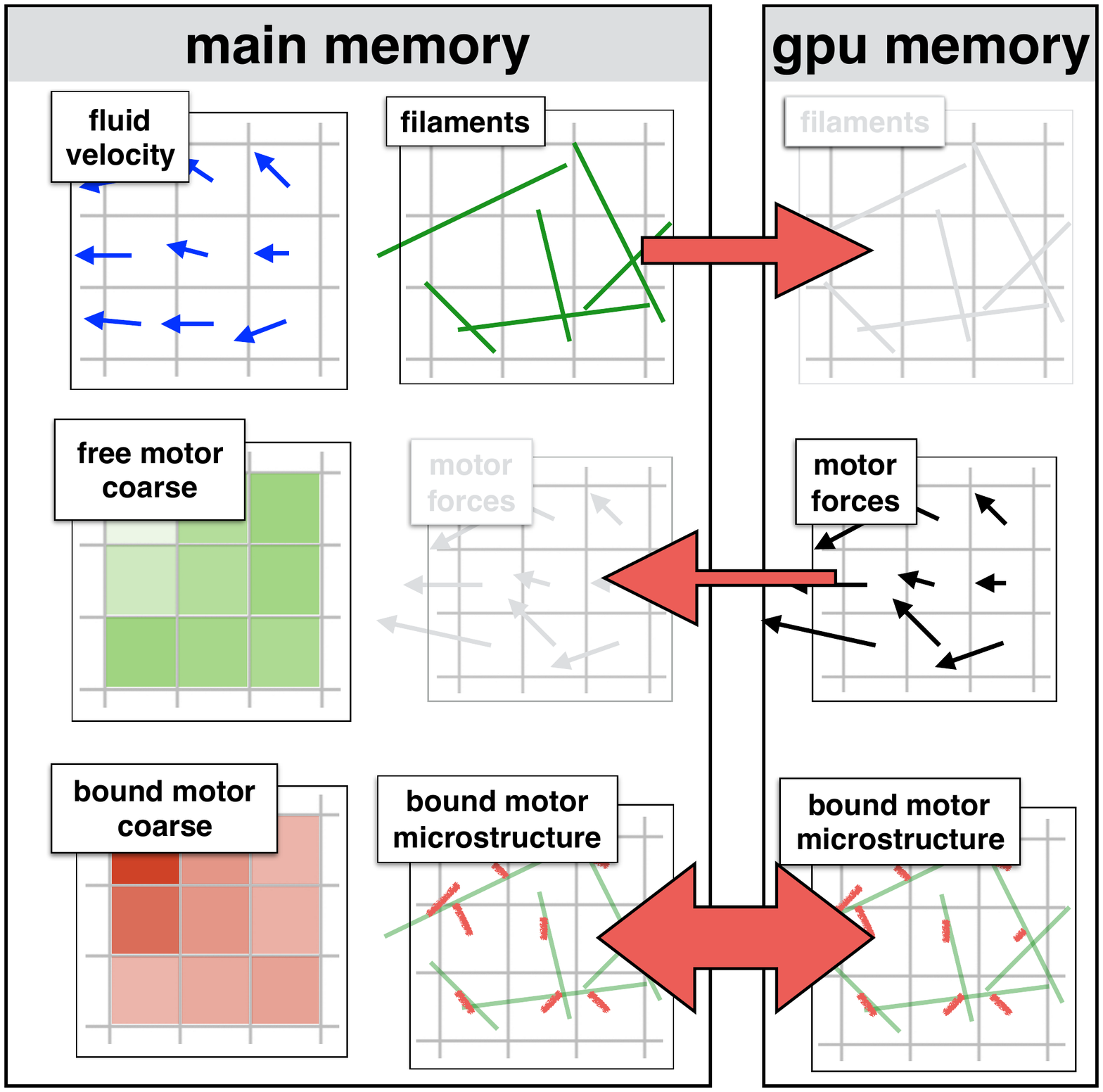}
\includegraphics[scale=.26]{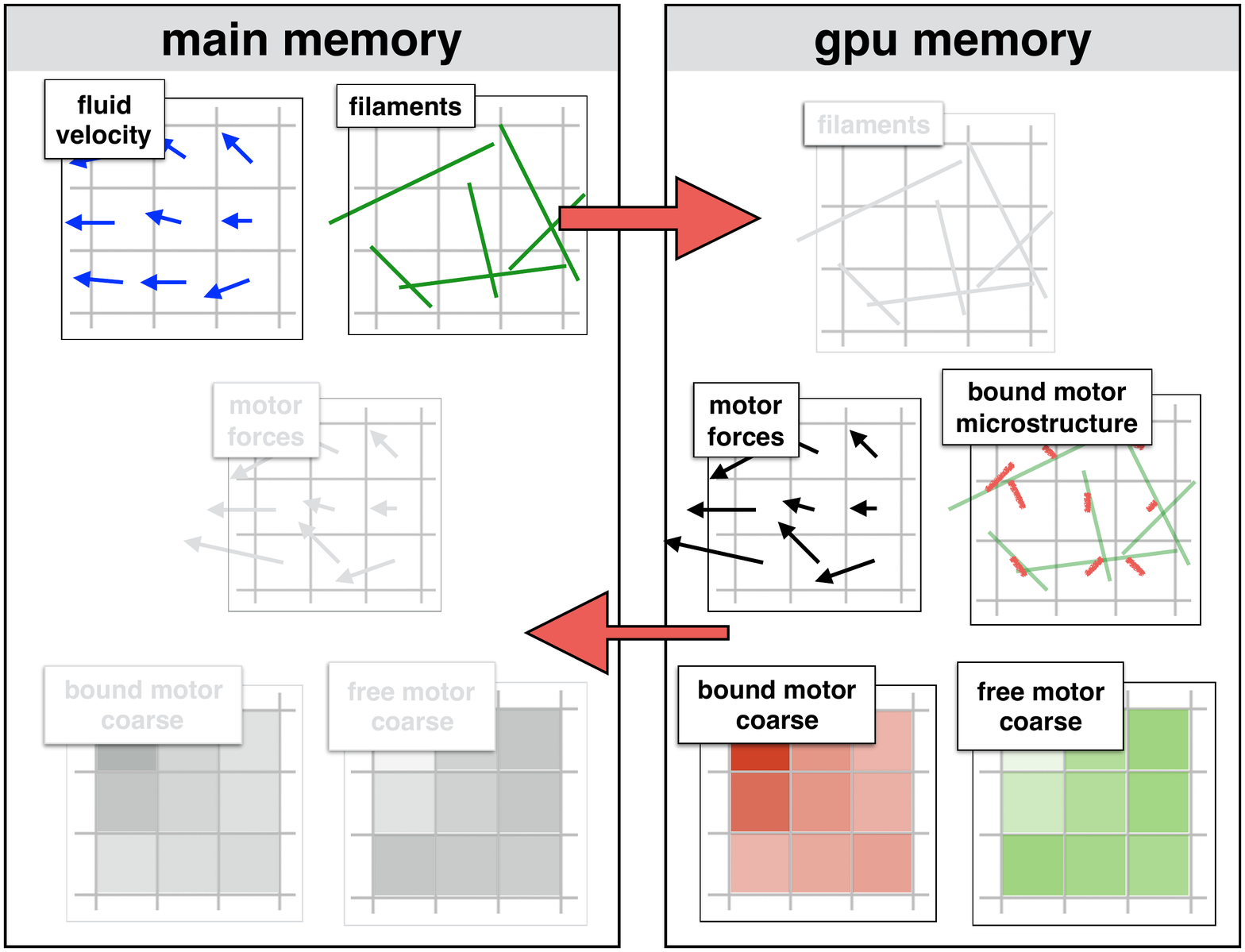}
\caption{Data residency and computation before (left) and after (right) restructuring. Left, transfer of the microstructure creates significant communication overhead. Right, the bound motor microstructure now fully resides on the GPU eliminating expensive transfers, and the coarse bound and free motor densities are calculated on GPU and transferred back to main memory.}
\label{fig:holistic}
\end{figure*}

A primary goal of our approach is to restructure the algorithm so that the memory-intensive microscale motor protein data resides solely on the GPU, and only the smaller, coarsened data is transferred to/from main memory. A secondary goal is to support a multi-GPU decomposition. Additionally, we remove synchronization barriers and reduce GPU memory consumption by two-thirds. Figure~\ref{fig:holistic} summarizes the changes and details follow.

\textit{Independent Time Steps.} The original algorithm calculated a global $dt^*$ and updated all $\M_b$ configurations by this fixed time step to time $t+dt$, hindering performance in several ways. First, it artificially limits the inherently independent per-cell update operations, some of which may be able to complete in fewer steps as their local configuration and stability restrictions allow. Second, it requires an expensive reduction operation over the entire bound motor density microstructure. Third, if $\M_b$ is distributed over multiple GPUs as desired, the reduction creates an unnecessary synchronization barrier. We instead compute a local $dt^*$ for each $\r_0$ at the beginning of each inner time step and update $\M_b$ at each $\r_0$ asynchronously. The most significant benefit of this change is enabling the multi-GPU implementation. Stability and accuracy were not adversely affected.

\textit{Numerical Integration Scheme.} The algorithm in \cite{CHS2017} used Adams-Bashforth 2 for time integration of the motor densities, which maintains the $\M_b$ array at three distinct time points ($t^{n+1},t^{n},t^{n-1}$). We instead use Runge-Kutta 2, which only requires the $\M_b$ array at $t^{n+1}$ and $t^{n}$. This change reduces GPU memory requirements by one-third while causing negligible impact on computation time. With these improvements, higher-resolution $\M_b$ density representations may reside in scarce GPU memory.

\textit{Mixed Precision.} We developed a mixed precision approach whereby we store and update $\M_b$ and $\M_f$ in single precision floating point while keeping the rest of the simulation as double precision. This saves space and improves performance without causing appreciable change in simulation behavior.

\begin{figure}
\centering
\includegraphics[scale=.2]{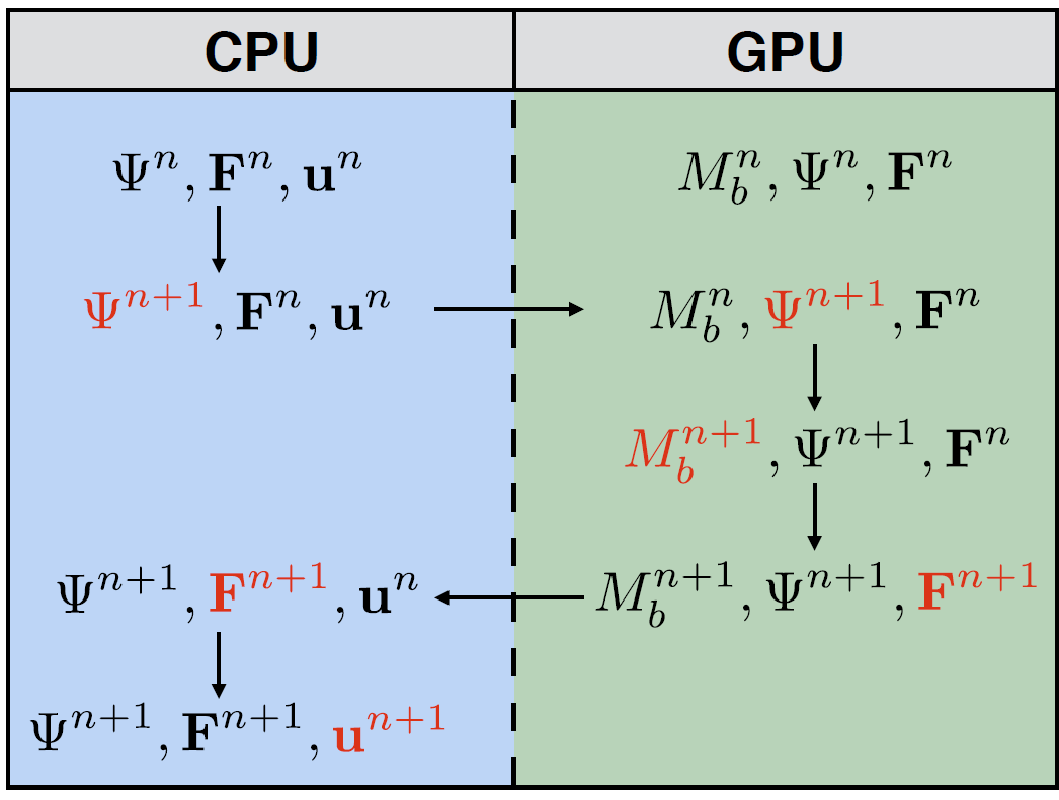}
\caption{Residency and evolution of state from time step $n$ to time step $n+1$ on CPU and GPUs in new algorithm. Red indicates the quantity updated through computation or data transfer.}
\label{fig:cpu_gpu_data_transfer}
\vspace{-.15in}
\end{figure}

\textit{Data Residency.}
The algorithm in \cite{CHS2017} updated $\M_b$ one piece at a time due to GPU memory constraints, then transferred the complete updated $\M_b$ to the GPU for the motor force calculation.  With the new numerical integration scheme and the use of mixed precision, we have enough GPU memory to store the high-dimensional microstructure data $\M_b$  solely on the GPU. This eliminates the overhead of transferring copies of $\M_b$ before, during, and after the $\M_b$ update. Since $\M_b$ is required in order to calculate $\M_{b,coarse}$ and $\vec{F}$, we also do those calculations on the GPU, and transfer results to the CPU. $\M_{b,coarse}$ and $\vec{F}$ are both macroscale data structures, and hence incur lower communication overhead. Finally, $\Psi$, which is also stored on the macroscale, is transferred as before.  The updated data flow is shown in Figure~\ref{fig:cpu_gpu_data_transfer}.

Pseudocode describing the new GPU kernel is presented in Algorithm~\ref{alg:evolution}. The result is one large kernel that fully updates $\M_b$ at each independent $\r_0$ value to $t+dt$ in as many steps as needed, using a local adaptive time step. The new memory access pattern is more amenable to caching as well, as each running block of threads on each GPU reads the same contiguous memory for all $\M_b$ configurations at a fixed $\r_0$ location repeatedly until those configurations are fully updated before moving on.

We use CPU parallelism via OpenMP to further accelerate the simulation, specifically in the calculations of the fluxes and stress tensors for the fluid solves, the outer global time step $dt$ calculation, and construction of the $\Psi$ advection matrix. After moving the $dt^*$ calculation into the motor force update kernel and multithreading the $dt$ calculation, time step calculation becomes a negligible component of the total computation time.

\subsection{GPU Optimizations}
This section describes various optimizations of the GPU kernel shown in Algorithm~\ref{alg:evolution}. Combined, these single-GPU optimizations yield an average improvement of 4.7-7.5x depending on resolution. The optimizations are described below and the individual effect of each is listed in Table~\ref{tab:gpu_optimizations}.

\textit{Mixed precision.} As previously detailed, switching $\M_b$, $\M_{b,coarse}$, and $\M_f$ to single precision halves the GPU memory requirement. In addition, it provides a 4.3x to 5.8x speedup in our bound motor density evolution routine. This improvement will depend on the clock cycle ratio between single and double precision arithmetic for a given GPU family.

\textit{Fast Math.} Compiling with CUDA's fast math library provides additional savings without noticeable change in simulation behavior. Accelerations of 1.35x were typical.

\textit{Launch Bounds.} The \textit{launch\_bounds} macro in CUDA may be used to instruct the compiler to ensure a user-specified maximum number of concurrent threads and threads per block running on each GPU Streaming Multiprocessor (SM). Using a \textit{launch\_bounds} configuration of 128 threads/block and 8 simultaneous blocks per SM gives the best performance of all configurations tested. Register spilling to global memory does occur at this configuration as each thread is limited to 32 registers. Newer architectures with more registers per SM will likely see immediate improvement by both reducing register spilling and enabling more threads per block. Accelerations of 1.2x were typical at the higher inner resolution and negligible at the lower inner resolution. 

\textit{Dimension Mapping.} CUDA threads are executed in simultaneous warps of 32 threads each, grouped first by their x-index then by their y-index. Since coalesced memory accesses are desirable for performance, the bound motor density evolution kernel was modified so that a thread's x-index maps to the $s$-index and the y-index maps to the $\theta$-index. With this mapping threads executing in a warp will access $\M_b$ storage in a coalesced fashion since sequential s-indices are contiguous as the innermost array indices. Accelerations of 1.16x to 1.57x were observed.

\textit{Reordering Storage.} The Superbee flux limiter operates in the arclength $s$ dimension. Since threads in a warp operate on subsequent arclength indices, and the flux limiter has a neighborhood access pattern of $(s-2, s-1, s, s+1, s+2)$, this gives coalesced memory accesses and pulls adjacent arclength data into the cache for subsequent iterations. The layout of memory in $\M_b$ was modified to make $s$ the innermost variable instead of $\theta$ in the storage of $\M_b(\r_0,\x,\theta,s)$, where $\x,\theta,s$ are represented as sequential flat four-dimensional arrays within a flat two-dimensional array over $\r_0$.  This prevents strides between subsequent $s$ accesses. Another benefit to making $s$ the innermost variable is that the value of $\Psi(\x,\theta)$ can be read after the $\theta$ loop instead of in the innermost loop. Reordering the loops in this fashion in the access-heavy motor force code resulted in a 1.54x  acceleration. For the bound motor density update, accelerations of 1.1x were typical at the higher inner resolution and negligible at the lower inner resolution.

\textit{Unrolling Reductions.} For the reduction step, \cite{harris2007optimizing} recommends manually unrolling a reduction when the number of remaining threads is less than the warp size (32 for our Tesla M2075), and performing part of a large reduction independently within each thread to reduce synchronization. We already follow the latter suggestion as each thread accumulates its contribution to $\M_{b,coarse}$ before storing this running sum in a shared memory array sized to the number of threads for the reduction step. We did not find meaningful performance improvements for the manual reduction.

\begin{algorithm}[t]
  \caption{Bound \& free motor density update GPU kernel \label{alg:evolution}}
    {Precondition: $\Psi(t+dt)$ and $dt$ are loaded into GPU memory.}\\
    {Set $t^*_{\text{end}}=t+dt$}\\
    \For{$\r_0 \in \text{grid}$}
    {
    \While{$t^* < t^*_{\text{end}}$}
    {
    {Compute adaptive $dt^*$}\\
    \For{$\x \in B_{rc}(\r_0)$}{
    \For{$\theta \in B_{rc}(\r_0)$}{
    \For{$s \in B_{rc}(\r_0)$}
    {         
    {Compute $\M_b(\r_0,\x,\theta,s,t^*+\frac{dt^*}{2})$ (Eq.~\eqref{eq:motor_1})} \\
    {Compute $\M_b(\r_0,\x,\theta,s,t^*+dt^*)$ (Eq.~\eqref{eq:motor_1})}\\
    }
    }
    }
    {Compute $\M_{\text{b,coarse}}(\r_0,t^*+dt^*)$ (Reduction) (Eq.~\eqref{eq:motor_2})}\\
    \If{$\M_{\text{total}}(\r_0)<\M_{\text{b,coarse}}(\r_0,t^*+dt^*)$}
    {
    \For{$\x, \theta, s \in B_{rc}(\r_0)$}
    {
    {Scale $\M_b(\r_0,\x,\theta,s,t^*+dt^*)$}\\
    {Set $\M_{\text{b,coarse}}(\r_0,t^*+dt^*)=\M_{\text{total}}(\r_0)$}
    }        
    }
    {Compute $\M_f(\r_0,t^*+dt^*)$ (Eq.~\eqref{eq:motor_3})}\\
    }   
    }
\end{algorithm}

\textit{Block Shaping.} To update $\M_b(\r_0,\x,\theta,s)$ we assign one block of threads to each $\r_0$ position and map those threads to the $\x,\theta,s$ variables. For each $\x$, a two-dimensional block of threads is launched, with the threads' x and y indices corresponding to the innermost $s$ and $\theta$ indices.  Experimentation has shown 128 threads to be the optimal number in our implementation. The ``block shaping'' row of the optimizations table compares against running 256 threads in a 16x16 configuration. How these 128 threads are configured is important: $x=8,y=16$ runs faster than $x=16,y=8$. Both caching effects and memory coalescing play a role, and from our experience it is worthwhile to experiment with various configurations. Accelerations of 1.15x were typical at the higher inner resolution and accelerations of 1.65x were typical at the lower inner resolution.

To analyze performance of the $\M_b$ and $\vec{F}$ GPU kernels, the Nvidia Visual Profiler v8.0 \cite{nvvp} was used. According to its output, arithmetic operations constitute the largest share of operations. No functional unit (load/store, arithmetic, control flow) is a bottleneck because of the balance of operations. We run the maximum possible number of simultaneous blocks per SM (8), but cannot run more threads per block without exhausting the available registers per SM. The result is a GPU occupancy of 66\%, for which the profiler's heuristics report that increasing occupancy is unlikely to improve execution time. Our experience confirms this, as attempts to launch more threads per block to increase occupancy means decreasing registers per thread to keep the simultaneous blocks per SM maximized at 8, resulting in longer execution times. GPU occupancy is one of many factors that contributes to kernel performance, and it is possible to obtain high throughput at low occupancy levels \cite{volkov2010better}.

\subsection{Scaling to Multiple GPUs}
As the spatial resolution of the $\r_0$ grid increases, two factors limit the performance of a single GPU. The first is that the number of blocks (each updating an independent $\r_0$) that can run concurrently on the GPU is limited by the number of SMs on the card, as we are running the maximum 8 simultaneous blocks per SM. Using two equivalent cards simultaneously doubles the throughput at which we can update the motor densities and calculate the motor force. The second factor is that once GPU memory is exhausted by the bound motor density (and scratch space for the intermediate values needed for numerical routines), additional large memory transfers to and from main memory become necessary every time step. 

\begin{figure}
\centering
\includegraphics[scale=.26]{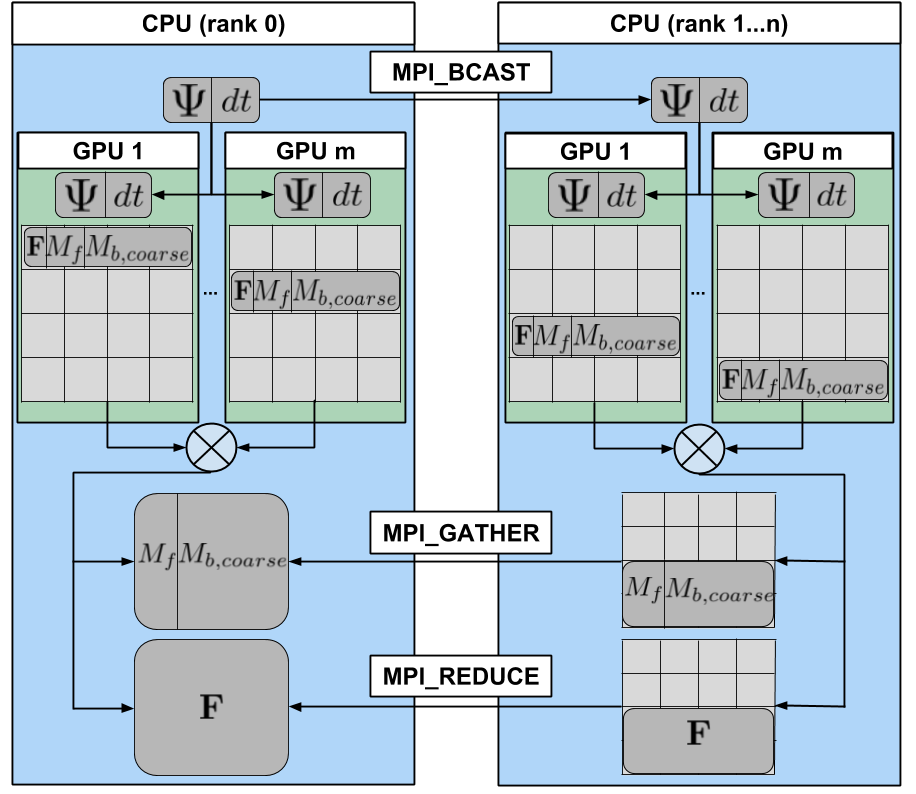}
\caption{MPI control flow for multiple GPUs across multiple nodes. We achieve nearly ideal scaling of our GPU computation across multiple GPUs, indicating that the MPI overhead is negligible. }
\label{fig:mpi}
\vspace{-.2in}
\end{figure}

We expand our implementation to multiple GPUs using simultaneous CUDA streams and to multiple machines using MPI. After $\M_b$ is initialized at the beginning of the simulation, subsections of $\M_b$ are transferred to the memory of each GPU. The outer two-dimensional array of $\M_b$ (over $\r_0$) is distributed among available GPUs by rows, which are contiguous in memory. Before invocation of the $\M_b$ and $\vec{F}$ kernels on the GPUs, the newly updated $\Psi$ is broadcast to each with an MPI\_BCAST from process rank 0. After the bound motor density update and motor force kernels complete, The $\M_f$ and $\M_{b,coarse}$ values computed by each GPU are collected by process rank 0 using an MPI\_GATHER operation. The motor force output $\vec{F}$ from each GPU contains overlapping force vectors that need to be summed together, so an MPI\_REDUCE operation is used to combine them in process rank 0. From here on the fluid velocity update proceeds as normal. The process is summarized in Figure~\ref{fig:mpi}. OpenMPI 2.1 was used for this work.

\section{Results}
\label{sec:results}

We present results at different resolutions, scaling both the outer resolution of the $x,y$ variables and the inner resolution of the $\theta,s$ variables. Increasing the outer resolution ($x,y$) affects the two-dimensional grid over which $\Psi,\M_f,\M_b$, and $\vec{u}$ are defined, and thus increases the workload across all steps of the simulation. Doubling the resolution of $x$ and $y$ increases by a factor of four the total workload of the $\M_b$ and $\vec{F}$ kernels. Increasing the inner resolution ($\theta,s$) affects $\Psi$ and $\M_b$ through their dependence on $\theta$ and $\M_b$ through its dependence on $s$. Doubling the resolution of $\theta$ and $s$ increases by a factor of four the number of update tasks per thread in the motor density and force kernels, and adds two more iterations to the reduction step in the motor density kernel.

Simulations were run on one or more servers configured with 2 Tesla M2075 GPUs, 64 GB RAM, and dual AMD Opteron 6272 processors. Speedup factors for the algorithmic modifications and GPU optimizations vs.\ the original implementation in a single-node, single-GPU configuration are shown in Table~\ref{table:single_node_scaling}. An average performance increase of between 5.75x and 9.98x per full simulation step is observed versus the original implementation. The bulk of the improvement comes in the bound motor update and motor force computations. Additionally, the $dt$ computation is accelerated, the $dt^*$ computation is moved onto the GPU, and the $\Psi$ solve noticeably benefits from CPU acceleration. 

\begin{table*}
  \begin{tabular}{ccccccl}
    \noalign{\hrule height 1.5pt}
    outer $\times$ inner resolution &$64^2\times16^2$&$64^2\times32^2$&$128^2\times16^2$&$128^2\times32^2$&$256^2\times16^2$&$256^2\times32^2$\\
    \hline
    total & 332 \textbf{(7.16)} & 854 \textbf{(8.76)} & 2054 \textbf{(6.37)} & 5026 \textbf{(9.55)} & 8980 \textbf{(7.36)} & 21408 \textbf{(9.25)} \\
    \hline
    $\Psi$ solve & 24 \textbf{(6.17)} & 49 \textbf{(5.37)} & 88 \textbf{(6.44)} & 171 \textbf{(7.88)} & 389 \textbf{(7.28)} & 722 \textbf{(9.19)} \\
    \hline
    $\M_b$ and $\vec{F}$ & 193 \textbf{(6.26)} & 682 \textbf{(6.42)} & 1492 \textbf{(6.03)} & 4366 \textbf{(7.51)} & 7050 \textbf{(6.65)} & 19125 \textbf{(4.72)} \\
    \hline
    fluid solve & 107 \textbf{(2.50)} & 112 \textbf{(2.21)} & 447 \textbf{(1.63)} & 456 \textbf{(1.96)} & 1437 \textbf{(2.00)} & 1438 \textbf{(2.42)} \\
    \noalign{\hrule height 1.5pt}
\end{tabular}
\caption{Effect of single-node, single-GPU optimizations. Average time per full outer time step (milliseconds) and speedup factors (bold) for original simulation vs.\ optimized simulation on a single machine with one GPU. Simulations with varied resolutions run to a fixed end time. The expensive $dt^*$ calculation in the original simulation is now negligible, contributing to the increased overall speedup reported in the total time step row.}
\label{table:single_node_scaling}
\end{table*}

We individually disable each GPU optimization and compare the running time for a single invocation of the $\M_b$ update kernel in Table~\ref{tab:gpu_optimizations}. We see the largest performance improvement from switching from double precision to single precision, which affects both floating point arithmetic performance as well as cache and memory demand. The $256^2 \times 32^2$ resolution could not be tested with double precision on a single GPU as $\M_b$ exceeded GPU memory. The launch\_bounds and storage reordering optimizations see their biggest impact when the inner variable resolution is increased. When this optimization was originally applied it showed a small improvement, but when it alone is removed from the final implementation, no discernible impact is observed.

\begin{table*}
  \begin{tabular}{lrrrrrr}
    \noalign{\hrule height 1.5pt}
    outer $\times$ inner resolution &$64^2\times16^2$&$64^2\times32^2$&$128^2\times16^2$&$128^2\times32^2$&$256^2\times16^2$&$256^2\times32^2$\\
    \hline   
    optimization removed: & & & & & & \\
    \hline
    none  & 350 & 710 & 1490 & 3560 & 5960 & 15660 \\
    \hline
    mixed precision & 1520 \textbf{(4.34)} & 4140 \textbf{(5.83)} & 7390 \textbf{(4.96)} & 19190 \textbf{(5.39)} & 30170 \textbf{(5.06)} & X \textbf{(X)}\\
    \hline
    -fast-math & 470 \textbf{(1.34)} & 870 \textbf{(1.23)} & 2010 \textbf{(1.35)} & 4530 \textbf{(1.27)} & 8050 \textbf{(1.35)} & 19710 \textbf{(1.26)}\\
    \hline
    launch\_bounds & 340 \textbf{(1.97)} & 860 \textbf{(1.21)} & 1490 \textbf{(1)} & 4270 \textbf{(1.2)} & 5890 \textbf{(99)} & 18200 \textbf{(1.16)}\\
    \hline
    dimension mapping & 440 \textbf{(1.26)} & 1120 \textbf{(1.58)} & 1820 \textbf{(1.22)} & 5210 \textbf{(1.46)} & 8200 \textbf{(1.38)} & 22420 \textbf{(1.43)} \\
    \hline
    reorder storage & 350 \textbf{(1)} & 820 \textbf{(1.15)} & 1490 \textbf{(1)} & 3960 \textbf{(1.11)} & 6010 \textbf{(1.01)} & 16950 \textbf{(1.08)} \\
    \hline  
    unroll reduction & 370 \textbf{(1.06)} & 700 \textbf{(0.99)} & 1510 \textbf{(1.01)} & 3630 \textbf{(1.02)} & 5920 \textbf{(0.99)} & 15350 \textbf{(0.98)} \\
    \hline
    block shaping & 550 \textbf{(1.57)} & 820 \textbf{(1.15)} & 2460 \textbf{(1.65)} & 4330 \textbf{(1.22)} & 9800 \textbf{(1.64)} & 18070 \textbf{(1.15)} \\
    \noalign{\hrule height 1.5pt}  
\end{tabular}
\caption{GPU optimizations. Time (milliseconds) and slowdown factors (bold) for the $\M_b$ evolution kernel at different resolutions with various optimizations individually disabled.}
\label{tab:gpu_optimizations}   
\end{table*}

\begin{figure}
\centering
\includegraphics[scale=.5]{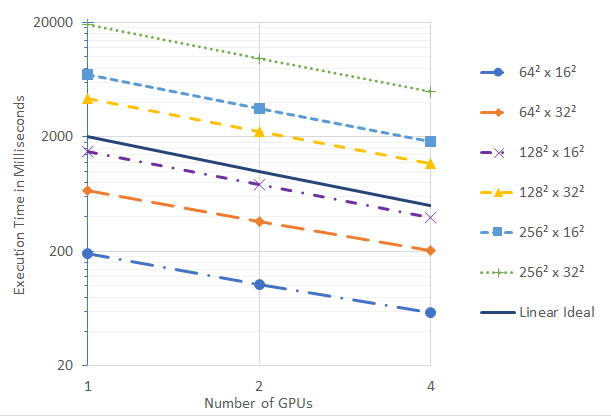}
\caption{Multi-GPU scaling of the bound motor density and motor force computations is nearly ideal. The overhead of inter-node communication via MPI arising in the 4 GPU configuration does not have an appreciable effect.}
\label{fig:gpu_scaling}
\end{figure}

Figure~\ref{fig:gpu_scaling} shows that scaling the $\M_b$ and $\vec{F}$ kernels to multiple GPUs and across nodes is effective, with nearly ideal linear acceleration at the higher resolutions where acceleration is most needed. This demonstrates that the overhead to merge output between GPUs on the same machine plus the MPI overhead among multiple machines is small compared to execution time. As the inter-node communication consists of MPI broadcast, reduction, and gather operations, it is expected that scaling to 8 or more GPUs would likewise involve minimal overhead cost.

\begin{figure*}
\includegraphics[scale=.14]{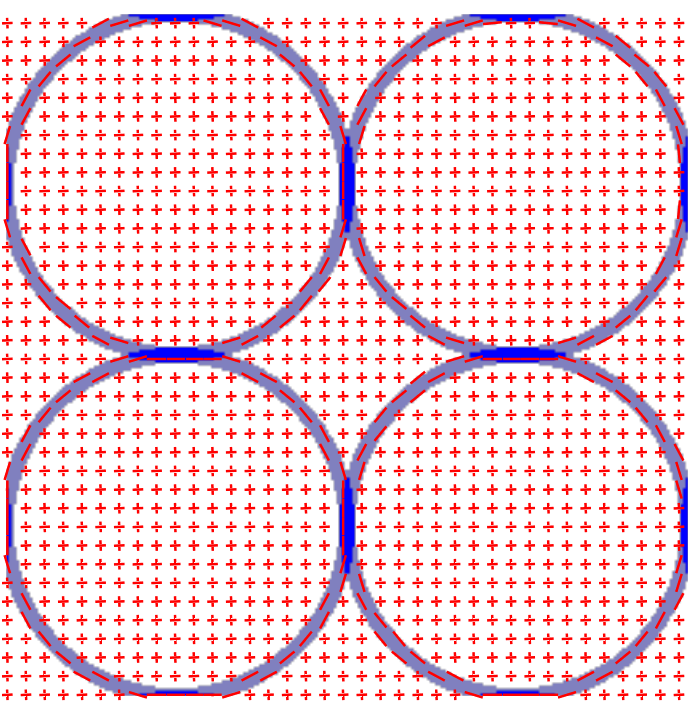}\includegraphics[scale=.14]{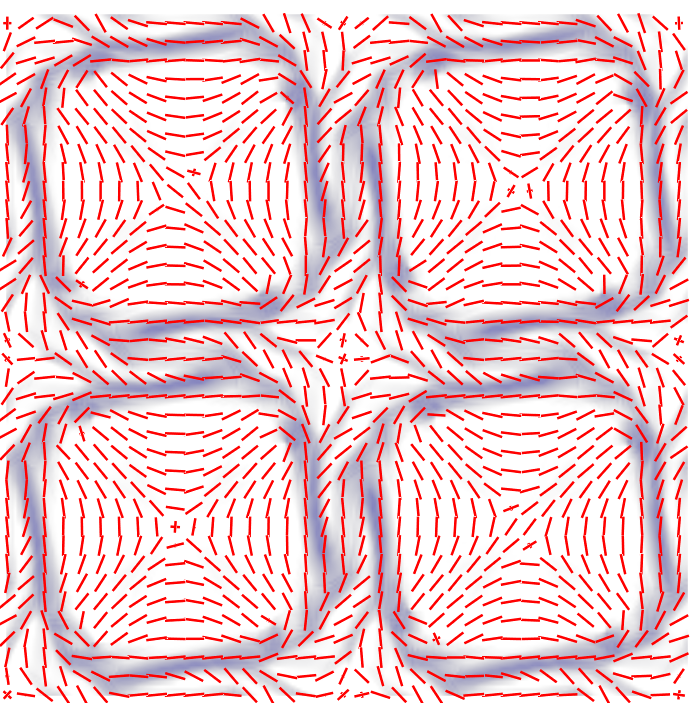}\includegraphics[scale=.14]{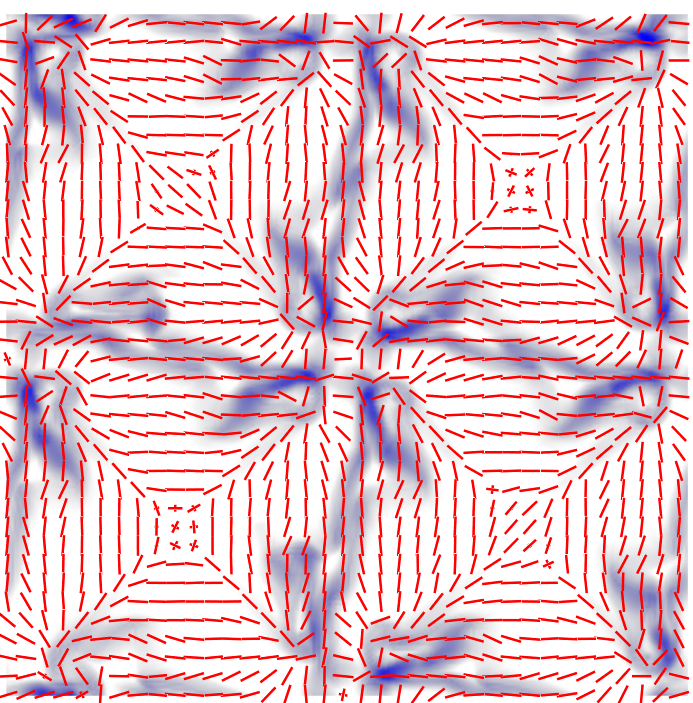}\includegraphics[scale=.14]{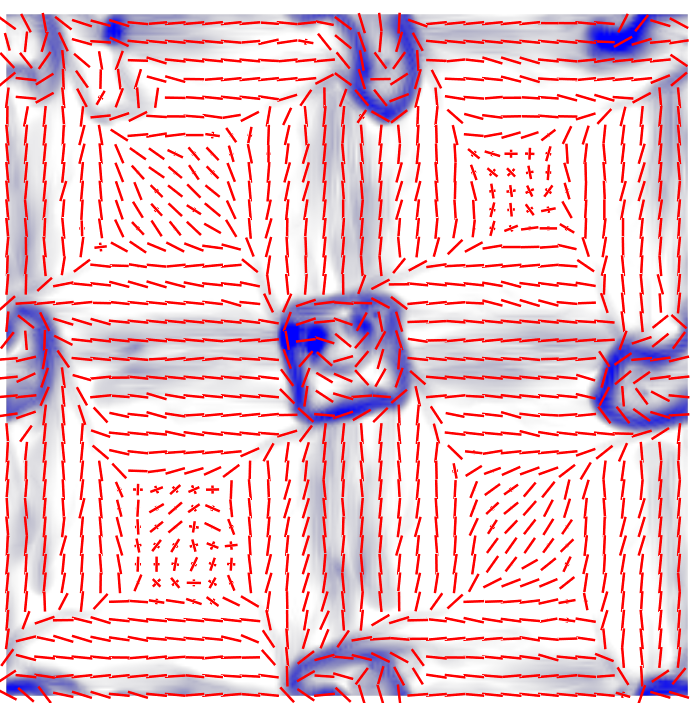}\includegraphics[scale=.14]{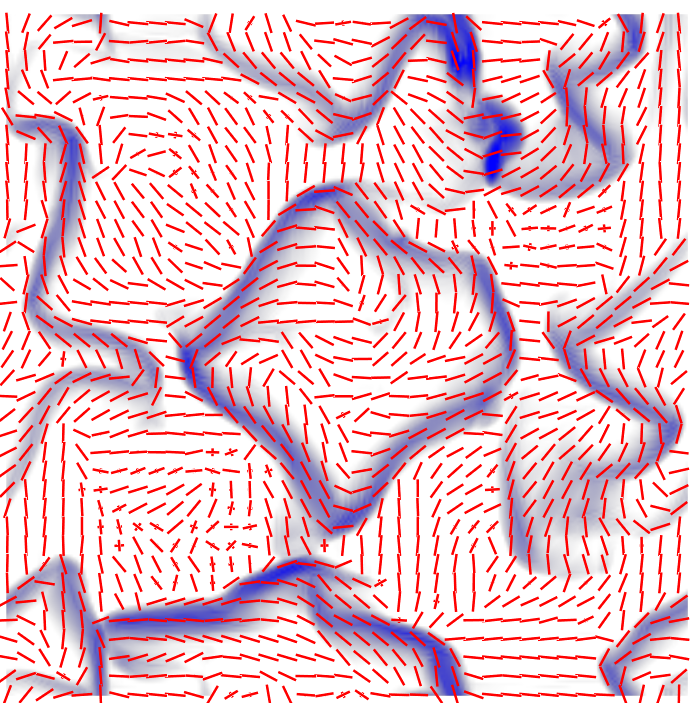}\\
\includegraphics[scale=.14]{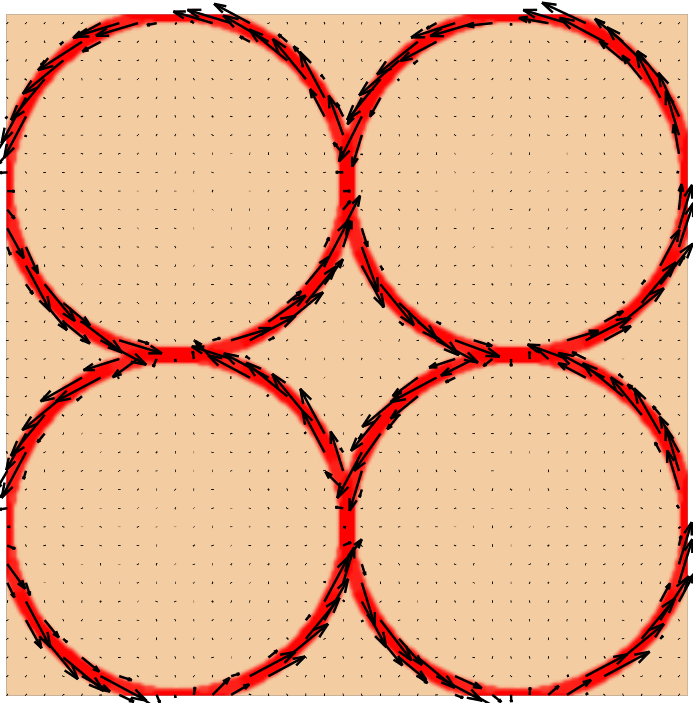}\includegraphics[scale=.14]{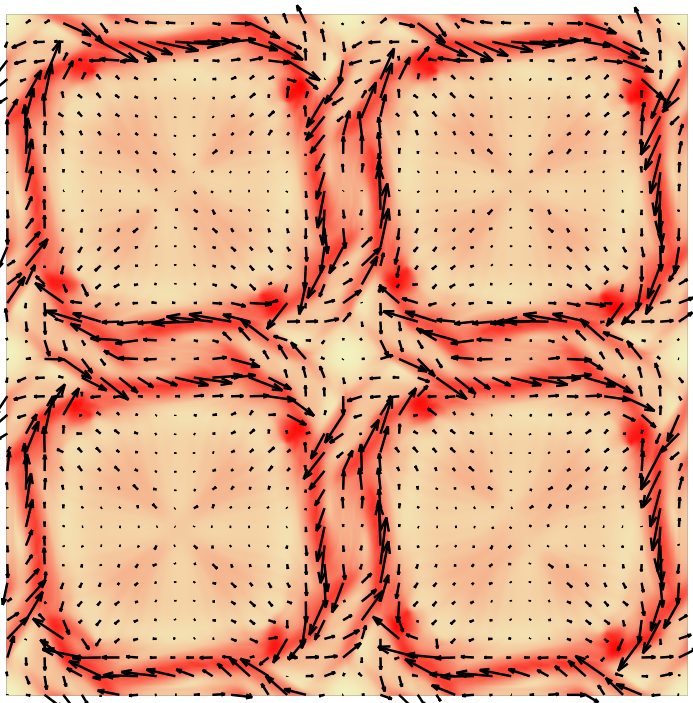}\includegraphics[scale=.14]{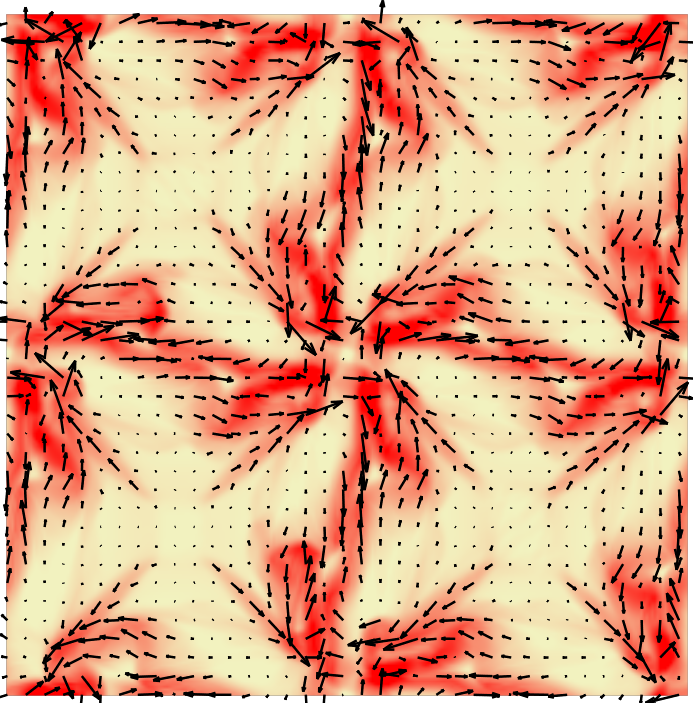}\includegraphics[scale=.14]{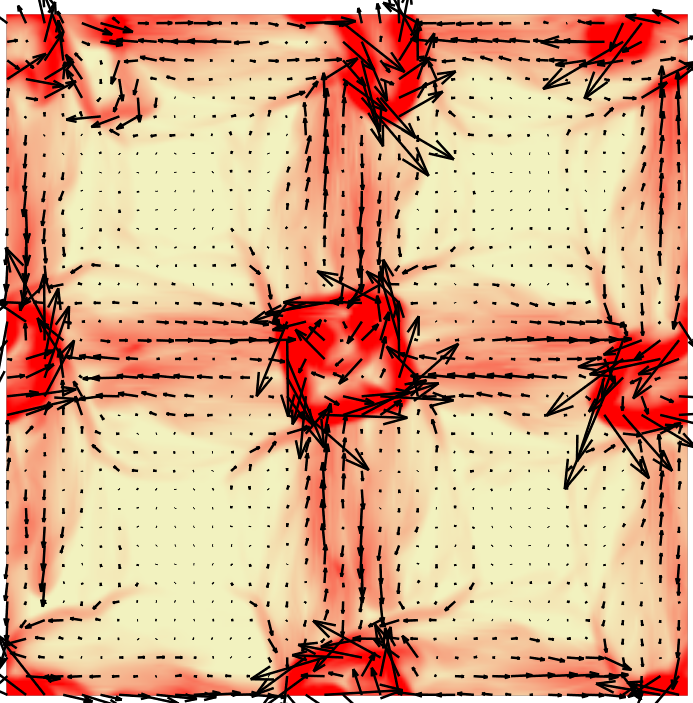}\includegraphics[scale=.14]{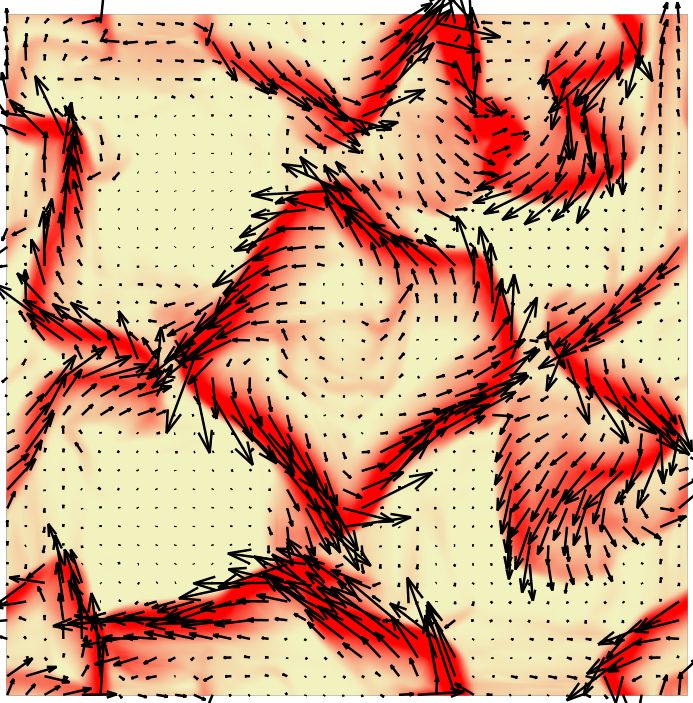}
\caption{Sample simulation output at equally spaced timing intervals of the evolution of a lattice of overlapping filament rings from an overhead view of the assay, with periodic boundary conditions. Top row: filament density $\Psi$ in blue with predominant orientation vector plotted every 8th cell in red. Bottom row: coarse bound motor density plotted in red with fluid velocity plotted as a black arrow every 8th cell.}
\label{fig:simulation_visualization}
\end{figure*}

Table~\ref{tab:multi_node_total_speedup} summarizes overall performance of our optimized implementation using one, two, and four GPUs vs.\ the original single-GPU only implementation. We obtain higher accelerations for higher inner resolution sizes, which is desirable as we find the inner resolution of $16^2$ too coarse at outer resolutions over $64^2$. Our maximum speedup factor over the original implementation was over 27x, obtained at the highest resolution. The simulation was previously limited to the $128^2 \times 32^2$ configuration given the running times involved. Reducing a day's worth of computation to less than one hour  greatly facilitates the iterative exploration of the model's parameter space. Sample simulation results at the previously infeasible $256^2 \times 32^2$ resolution are shown in Figure~\ref{fig:simulation_visualization}.

\begin{table*}
  \begin{tabular}{ccccccl}
    \noalign{\hrule height 1.5pt}
    outer $\times$ inner resolution &$64^2\times16^2$&$64^2\times32^2$&$128^2\times16^2$&$128^2\times32^2$&$256^2\times16^2$&$256^2\times32^2$\\
    \hline
    1 node,1 GPU & 7.16 & 8.76 & 6.37 & 9.55 & 7.36 & 9.25 \\
    \hline
    1 node, 2 GPU & 9.70 & 14.04 & 9.73 & 16.50 & 11.98 & 16.60 \\
    \hline
    2 node, 4 GPU & 11.98 & 20.23 & 13.59 & 26.69 & 17.69 & 27.41 \\
    \noalign{\hrule height 1.5pt}
\end{tabular}
\caption{Overall speedup resulting from our optimized, multi-GPU approach, as compared with original single-GPU simulation, as the number of nodes \& GPUs is increased.}
  \label{tab:multi_node_total_speedup}
\end{table*}

\section{Conclusions}
\label{sec:conclusions}

Mapping the different scales of a simulation to different computational hardware, minimizing data transfers, and removing synchronization points like a global time step calculation allows us to explore the parameter space of our high-dimensional micro-macro simulation up to 540 times faster than a single-threaded implementation when using four GPUs simultaneously. This holistic approach significantly outperforms the commonly employed approach of accelerating individual functions in isolation \cite{HCS2014}. Multi-node, multi-GPU overhead is minimal and the approach is expected to scale well to a greater number of GPU accelerators. This approach capitalizes on the increasing prevalence of GPUs in high performance computing.

As the number of GPUs used increases the semi-spectral fluid solve and filament evolution update steps will become the next bottlenecks. Further adjustment of simulation flow to offload more of the fluid solve computations onto the otherwise idle CPU cores of non-root processes may then become cost effective.

It is our hope that our algorithmic design and breakdown of the various CPU and GPU optimizations will provide a useful reference for prioritizing optimizations in HPC software development and in porting of existing applications, where there is often an expectation that porting time should be recovered by faster runtimes. While the specific improvement will vary for different programs, quantifying the improvements corresponding to various optimizations contributes to the growing information in the literature regarding their efficacy \cite{niemeyer2014recent}.


%% file: commands.tex
\renewcommand{\vec}[1]{\mathbf{#1}}
\newcommand{\symb}[1]{\boldsymbol{#1}}
\newcommand{\x}{\vec{x}}
\newcommand{\X}{\vec{X}}
\newcommand{\p}{\vec{p}}
\newcommand{\y}{\vec{y}}
\newcommand{\z}{\vec{z}}
\newcommand{\I}{\vec{I}}
\renewcommand{\r}{\vec{r}}
\newcommand{\gon}{\gamma^{+}}
\newcommand{\goff}{\gamma^{-}}
\newcommand{\uvec}{\vec{u}}
\renewcommand{\L}{\mathcal{L}}
\newcommand{\M}{\mathcal{M}}
\newcommand{\Lt}{\mathcal{L}^{\star}}
\newcommand{\Ltp}{{\mathcal{L}'}^{\star}}
\newcommand{\teps}{\varepsilon}
\newcommand{\deltaeps}{\delta_{\teps}}
\newcommand{\pperp}{(\vec{I}-\p\p)}
\newcommand{\intl}{\int_{-l}^l}
\newcommand{\bars}{\bar{s}}
\newcommand{\uinf}{\vec{u}_{\infty}}
\newcommand{\uinfs}[1]{\vec{u}_{\infty}(\X+#1\p)}
\renewcommand{\S}{\boldsymbol{\sigma}}
\newcommand{\Sp}{\S^\text{p}}
\newcommand{\Sf}{\S^\text{f}}
\newcommand{\St}{\S^\text{t}}
\newcommand{\Sfp}{\S^\text{f'}}
\newcommand{\Stp}{\S^\text{t'}}
\newcommand{\Ssp}{\S^\text{s'}}
\newcommand{\Spp}{{\S'}^{(p)}}
\newcommand{\Dx}{D_t}
\newcommand{\Dxpar}{D_{\textrm{\upshape{t}},||}}
\newcommand{\Dxperp}{D_{\textrm{\upshape{t}},\perp}}
\newcommand{\Dp}{D_\textrm{\upshape{r}}}
\newcommand{\micron}{\mu\textrm{\upshape{m}}}
\newcommand{\millimeter}{\textrm{\upshape{mm}}}
\newcommand{\nanometer}{\textrm{\upshape{nm}}}
\newcommand{\micromolar}{\mu\textrm{\upshape{M}}}
\newcommand{\second}{\textrm{\upshape{s}}}
\newcommand{\piconewton}{\textrm{\upshape{pN}}}
\newcommand{\fperp}{\vec{f}_{\perp}}
\newcommand{\thetahat}{\hat{\boldsymbol{\theta}}}
\newcommand{\phihat}{\hat{\boldsymbol{\phi}}}
\newcommand{\mylistone}[1]{\begin{tabular}{l} {#1} \end{tabular}}
\newcommand{\mylisttwoh}[2]{\begin{tabular}{ll} {#1} & {#2} \end{tabular}}
\newcommand{\mylisttwov}[2]{\begin{tabular}{l} \\{#1}\\{#2} \end{tabular}}
\newcommand{\mylistthreev}[3]{\begin{tabular}{l} {#1}\\[2ex]{#2}\\[2ex]#3 \end{tabular}}
\newcommand{\mylistfourv}[4]{\begin{tabular}{l} \\{#1}\\[2ex]{#2}\\[2ex]{#3}\\[2ex]{#4}\\[2ex] \end{tabular}}
\newcommand{\mylistfivev}[5]{\begin{tabular}{l} \\{#1}\\[2ex]{#2}\\[2ex]{#3}\\[2ex]{#4}\\[2ex]{#5}\\[2ex] \end{tabular}}
\newcommand{\mylistsixv}[6]{\begin{tabular}{l} \\{#1}\\[2ex]{#2}\\[2ex]{#3}\\[2ex]{#4}\\[2ex]{#5}\\[2ex]{#6}\\[2ex] \end{tabular}}
\renewcommand{\odot}[1]{\mathring{#1}}

\definecolor{todoColor}{rgb}{0.8,0,0.8}
\newcommand{\todo}[1]{{#1}}